\documentclass[9.5pt, conference, twocolumn]{IEEEtran}
\usepackage{graphicx, hyperref,epsfig}
\usepackage{eurosym}
\usepackage{subfigure}
\usepackage{url}

\begin{document}

\title{An e-Infrastructure for Collaborative Research in Human Embryo Development}

\author{\IEEEauthorblockN{Adam Barker\IEEEauthorrefmark{1},  Jano I. van Hemert\IEEEauthorrefmark{2}, Richard A. Baldock\IEEEauthorrefmark{3} and Malcolm P. Atkinson\IEEEauthorrefmark{2}}. 

\IEEEauthorblockA{\IEEEauthorrefmark{1}Department of Engineering Science, University of Oxford, UK.}  
\IEEEauthorblockA{\IEEEauthorrefmark{2}National e-Science Centre, School of Informatics, University of Edinburgh, UK.}
\IEEEauthorblockA{\IEEEauthorrefmark{3}Human Genetics Unit, Medical Research Council. UK}}

\maketitle

\begin{abstract} 

Within the context of the EU Design Study \emph{Developmental Gene Expression Map}, we identify a set of challenges when facilitating collaborative research on early human embryo development. These challenges bring forth requirements, for which we have identified solutions and technology. We summarise our solutions and demonstrate how they integrate to form an e-infrastructure to support collaborative research in this area of developmental biology. 
\end{abstract}


\section{Introduction} \label{s:introduction}

Characterising gene expression patterns is a crucial part of understanding the molecular determinants of embryonic development and the role of genes in disease. However, gene expression studies in human embryos need to overcome a number of difficulties. These include the sourcing and maintenance of collections of human material suitable for gene expression studies, bridging the expertise in both biological and informatics areas, and amassing and making accessible data from multiple laboratories and studies. 

Embryonic gene expression studies are common place in model organisms such as drosophila (fly), mus musculus (mouse) and danio rerio (zebra fish), where procedures such as \emph{in-situ} hybridisation allow scientists to detect and localise the presence of RNA transcriptomes that are likely to correspond to gene activity. This type of study is less common in humans, and currently only two laboratories are licensed in the UK to collect and analyse human embryonic tissue: the Human Developmental Biology Resource (HDBR) at Newcastle University and the MRC Fetal Tissue Bank (FTB) at University College London (UCL)~\cite{LindsayCopp2005}. Human embryonic material is extremely scarce world-wide, dissemination of the results to the wider community is therefore vital for the progression of research.

\emph{Developmental Gene Expression Map} (DGEMap) (\url{http://www.dgemap.org/}) is a EU-funded Design Study, which will accelerate an integrated European approach to gene expression in early human development. It runs from April 2005 until March 2009. Due to the scarce nature of human tissue, it is the first project of its kind in the world. DGEMap was granted \euro{2.2} million and involves collaborations between the Institute of Human Genetics at Newcastle University, and the National e-Science Centre at the University of Edinburgh. DGEMap comprises a number of work packages that span two major disciplines: biology and informatics. Gene expression studies on a small-scale are common place, the first work package investigates the difficulties of automating systematic gene expression studies on a large-scale. An ethics study evaluates the inconsistent world-wide regulations and legislation on the use of human embryos.  A feasibility study focuses on a user requirements of any resulting framework. The final work package and the focal point of this paper focuses on designing a suitable e-infrastructure to support distributed laboratories. This paper provides an overview of all involved solutions and technologies, it does not aim to give complete details of each solution; instead references to complimentary papers are provided throughout the text. 

\section{DGEMap Challenges}

Establishing and maintaining suitable collections of human developmental material requires time, significant resources and knowledge that a single laboratory will find difficult to sustain. These difficulties may be overcome by linking multiple laboratories and individual experts in the field, providing a single point of access. A framework that facilitates such collaborative research brings a number of key challenges:

\medskip

\textbullet~\textbf{Distributed laboratories} Researchers distributed over various physical locations around the world should be able to formulate and collaborate on research projects that involve wet-lab experiments on human embryo tissue. This requires dealing with providing secure access of researchers to selected subsets of several tissue banks, the experimental data and results. In the DGEMap project this is dealt with by a Web Portal, which is discussed in detail in Section~\ref{s:portal}. Moreover, ethical legislation differs over countries, which can in many locations make embryonic tissue unavailable, or even make experiments on human embryonic tissue illegal. In close collaboration with Policy, Ethics and Life Sciences at Newcastle University, the project has evaluated the status of regulations and legislation on the use of human developmental material and surveys professional standards~\cite{TaylorWoods2008}.

\medskip
\textbullet~\textbf{Distributed data handling} Wet-lab experiments will be performed at several locations, the results of which are mostly stored at the site where the experiments were performed. When researchers are querying, processing or analysing data these data need to be transferred to other locations, either for processing or for visualisation. To cope with the large quantity and highly complex research data any resulting e-infrastructure needs to provide transparent mechanisms to the underlying workflows as well as allow for large amounts of data to be moved efficiently between sites. Issues regarding the transportation and processing of large amounts of mostly image data are discussed in Section~\ref{s:workflow} and Section~\ref{s:imageprocessing}.  

\medskip
\textbullet~\textbf{Coping with a scarce resource} Human embryo tissue is a scarce resource, which makes finding techniques that maximise the results from each section of tissue a priority. One of the objectives of the DGEMap project is to optimise the physical use of tissue through new wet-lab techniques and more reliable instruments such as automatic systems for immunohistochemistry and in situ hybridisation. Also, after a small set of results from these instruments are captured as images, we can apply image processing techniques to accurately predict results. A detailed description of this process is provided in Section~\ref{s:imageprocessing}.

\medskip
\textbullet~\textbf{Few existing results on human development} Where human embryo tissue is a scarce resource, with as a result not much data and information about the developmental processing in terms of genetics; much is known about the development in, for example, mouse models. By linking the results from DGEMap to results in external data sources, we can aid researchers in finding directions for future studies. Details about the possibilities for linking on various levels is provided in Section~\ref{s:humanmouse}. When these links are provided as services, they can be used by researchers to compose workflows that extract and process data and information to go back and forth between the mouse and human embryo databases.

\medskip
\textbullet~\textbf{Large amounts of mouse-model results} When data from the human embryo is linked with data and information from the mouse, we end up with a very rich resource. The disadvantage is that it will become more difficult to extract relevant and significant ideas for future studies. It is vital for an e-infrastructure with access to large quantities of data to provide methods to enable hypothesis generation. In Section~\ref{s:datamining}, we discuss in detail two methodologies from the field of data mining that directly operate on the spatio-temporal nature of gene expression patterns.

\section{e-Infrastructure} 

\begin{figure} [htp]
\centering
\includegraphics[%
width=0.5\textwidth]{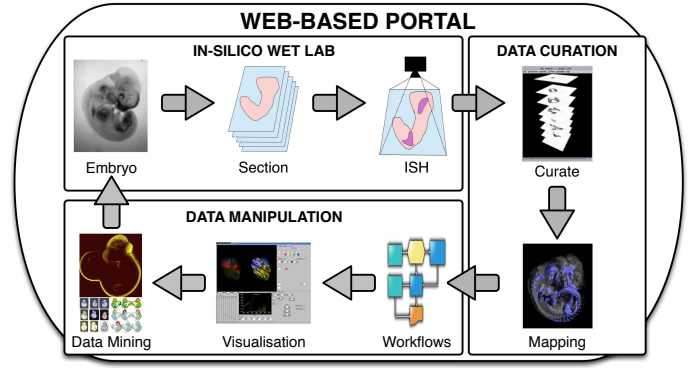}
\caption{Flow of tasks in a gene expression study from an embryo through to \emph{in-situ} experiments to mapping results to embryo models to data manipulation through workflow, visualisation and data mining techniques. The arched arrow illustrates how the results of data manipulation tasks lead to hypothesis generation which influences future wet-lab experimentation. All tasks are accessible via a Web browser through the DGEMap Web portal. Note that the images shown here are for illustrative purposes.}
\label{fig:study}
\end{figure}

The e-infrastructure must support the tasks involved in high resolution \emph{in situ} gene expression studies. In Figure~\ref{fig:study} we show the basic flow of tasks in such a typical study. First, an embryo is frozen or embedded in a block of wax after which it can be cut into sections, which are all placed on slides. These sections will undergo either \emph{in-situ} hybridisation or immunohistochemistry, which are histological techniques to reveal by colourmetric, radioactive or fluorescent stains the distribution of gene expression products. These products are either the transcription product mRNA or the resultant protein and are the evidence for gene-expression in any given cell. The sections are viewed using microscopy and the tissue structure and expression patterns digitally captured to images. The digital images are then aligned or mapped on to a 3D reference embryo model of the same developmental stage. Thereafter, we can query expression patterns from multiple genes, which allows us to do large scale analyses and data mining to extract knowledge and consequently synthetise new hypotheses.

Due to the scarce nature of human embryonic material it is essential to allow uniform access to these data to remote researchers. With this requirement in mind, a Web-based portal delivers a single entry point for EU research enabling collaborative \emph{in-situ} experiments on human embryo tissue. The federated approach allows researchers to query tissue banks in multiple distributed sites, at the same time. Workflows within the portal are data-centric, optimisation is therefore vital when moving large data for analysis. Image capturing techniques provide mechanisms to automatically digitise embryonic samples and capture relevant meta-data, allowing external scientists to search, analyse and verify these experimental data. After mapping 2D histological sections to standard 3D spatio-temporal models, data mining is performed to find ``interesting'' interaction patterns between sets of genes. The resulting DGEMap e-infrastructure provides a unified solution for human embryo research, all aspects of functionality are provided through a Web-based portal, making it possible to access data and results through a Web browser. In the following sections, we will discuss each component in detail.

\subsection{DGEMap Web Portal}\label{s:portal}

\begin{figure}[htp]
\centering
\includegraphics[%
width=0.50\textwidth]{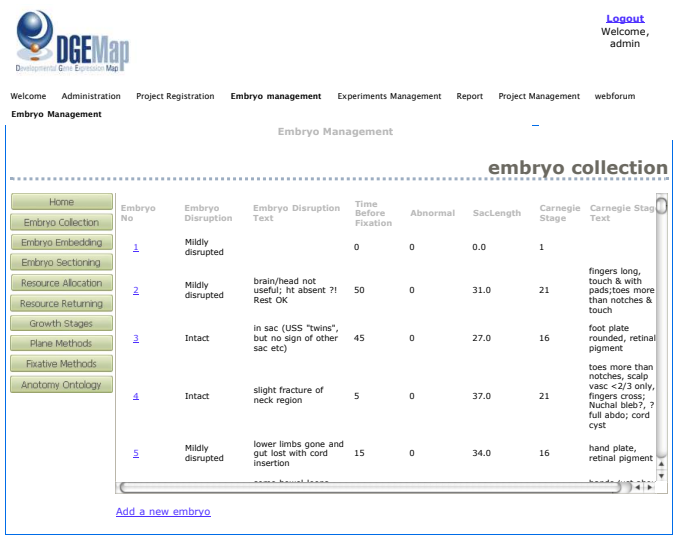}
\caption{A screenshot of the DGEMap Web Portal, showing the embryo collection available at the HDBR in Newcastle
\vspace{-1\baselineskip}}
\label{fig:portal}
\end{figure}

DGEMap provides a single point of access to multiple distributed gene expression databases, this is achieved through the use of \emph{portal} technology, in particular a portal implementation provided  through the open-source Gridsphere project ({\url{www.gridsphere.org/}). Portals provide access to groups of diverse Web sources and allow users to customise the way in which these data are viewed. Portal implementations provide a component model that allows autonomous components known as \textit{portlets} to be plugged into a portal infrastructure. Portlets represent modular, reusable software components that may be developed independently of the general portal architecture and offer a specific set of operations.

The DGEMap portal serves as a key resource for researchers involved at each stage in the gene expression study cycle detailed in Figure~\ref{fig:study}, in particular human embryonic gene expression studies. The portal captures the major tasks of HDBR managers, staff, and resource users, and provides facilities for project management, embryo search, resource distribution and tracking and efficient information sharing and manipulation, through workflow technology (discussed in Section~\ref{s:workflow}), image manipulation and visualisation (discussed in Section~\ref{s:imageprocessing}) and data mining techniques (discussed in Section~\ref{s:datamining}). 

\subsection{Workflow Optimisation - The \emph{Circulate} Architecture} \label{s:workflow}

Efficiently executing large-scale, data-intensive workflows~\cite{ppam08} common to scientific applications (i.e. scenarios detailed by DGEMap) must take into account the volume and pattern of communication. Standard workflow tools based on a centralised orchestration engine can easily become a performance bottleneck for such applications, extra copies of the data (\emph{intermediate data}) are sent that consume network bandwidth and overwhelm the central engine. In order to address this problem the DGEMap infrastructure proposes the \emph{Circulate} architecture, a novel solution based on centralised control flow, distributed data flow~\cite{service_models}. The \emph{Circulate} architecture sits between pure orchestration (completely centralised) and pure choreography (completely decentralised). A centralised orchestration engine issues control flow messages to services taking part in the workflow, however enrolled services can pass data flow messages amongst themselves, like in a peer-to-peer system. This model maintains the robustness and simplicity of centralised orchestration but facilities choreography by avoiding the need to pass large quantities of intermediate data through a centralised server.

\begin{figure*} [htp]
\centering
\includegraphics[%
width=0.7\textwidth]{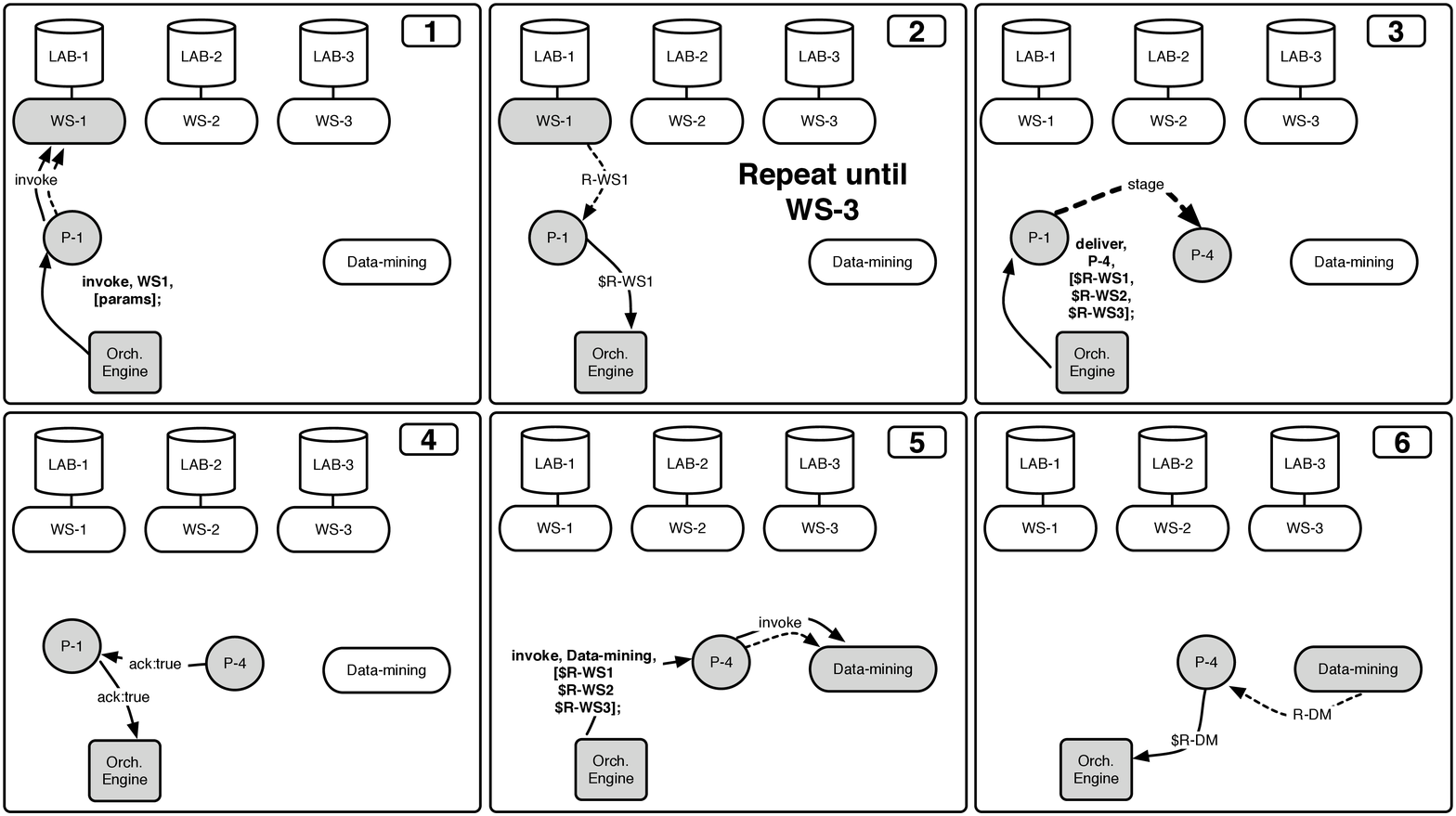}
\caption{Centralised control flow, distributed data flow workflow architecture}
\label{fig:dataflow}
\end{figure*}

The solution is best illustrated through an example, with reference to Figure~\ref{fig:dataflow}, a number of distributed laboratories conducting human embryonic gene expression studies offer data via a Web service interface. A scientist needs to retrieve and collate data from these three laboratories and run the resulting data set through a data mining algorithm. The first steps in the workflow (phases 1--2 in Figure~\ref{fig:dataflow}) involve retrieving data from the laboratories via Web service interfaces. However, instead of contacting the service directly, a call is made to a proxy (in the first instance \texttt{P-1}) which is installed as ``near" as possible to enrolled Web services; by near we mean preferably on the same Web server or network domain, so that communication between a proxy and a Web service takes place over a local network. The proxy invokes the service and the output, in this case \texttt{R-WS1} is passed back to the proxy, tagged with a unique identifier and stored within the proxy. Instead of the proxy returning these data to the orchestration engine a reference to the data is returned instead, in this case, \texttt{\$R-WS1}. This process is repeated for \texttt{WS-2} and \texttt{WS-3}, retrieving the necessary data from each laboratory. The output from the Web service invocations, \texttt{\$R-WS1, \$R-WS2, \$R-WS3} are needed as input to the \texttt{Data-mining} Web service (phases 3--4). The orchestration engine instructs \texttt{P-1} to move these data to a proxy sitting near the \texttt{Data-mining} Web service, \texttt{P-4}. These data are used as input to the \texttt{Data-mining} Web service via the proxy (phases 5--6), the output of which, \texttt{R-DM} is stored at the proxy, which in turn returns a reference to these data to the orchestration engine, \texttt{\$R-DM}.

The \emph{Circulate} architecture allows control flow messages to pass through the orchestration engine, while larger data flow messages (containing embryo images etc.) to be exchanged between proxies in a peer-to-peer fashion. Details of the API and implementation of the proxy can be found in a complementary paper~\cite{ccgrid08}. Performance analysis~\cite{hpdc08} concludes that a substantial reduction in communication overhead results in a 2--4 fold performance benefit across common workflow patterns, essential when dealing with the data sizes involved in DGEMap scenarios.

\subsection{Image Processing and Visualisation}\label{s:imageprocessing}

\textit{In situ} gene expression studies intrinsically make use of image output and image processing forms an important role in the e-infrastructure. To analyse the gene-expression data it is important to be able to compare expression domains from multiple genes that could be part of the same genetic network. To enable this comparison of several gene expression patterns, databases such as the Embryonic Atlas of the Developing Human Brain (EADHB) and the Edinburgh Mouse Atlas Project (EMAGE)~\cite{ChristiansenYang2006a} have developed techniques to spatially map data from section or 3D data directly onto standard reference models. The mapping process consists of image warping from the experimental embryo onto the reference model. With embryonic material the required spatial transformation can be very complex requiring significant expert input and image processing. These techniques are under active research, the challenge for the e-infrastructure is to provide the tools to the user, capture the resultant transformations and to transform the image data.

One of the major challenges is to deal with the scarce nature of the material. In the context of the e-infrastructure, this has lead to new methods for extrapolating results from sparsely sampled gene expression patterns. An example is shown in Figure~\ref{fig:pattern_extrapolation} where a small set of gene expression patterns have been mapped on to a 3D embryo model of the same developmental stage as the embryo from which sections were derived. After this, it is possible to predict the expression pattern between the two outer sections by extrapolation all the sections in between the area of the original sections.

\begin{figure}[htp]
\begin{center}
\subfigure[A set of original sections projected in a 3D reference embryo]{\includegraphics[width=0.475\textwidth]{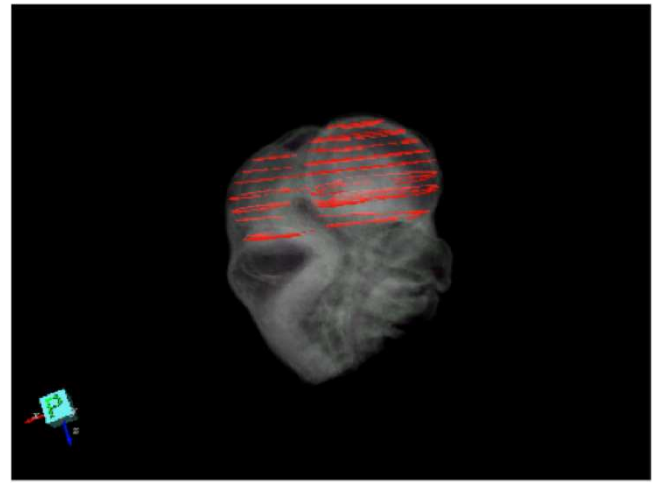}}
\hspace{1em}
\subfigure[The predicted expression pattern based on the original sections]{\includegraphics[width=0.475\textwidth]{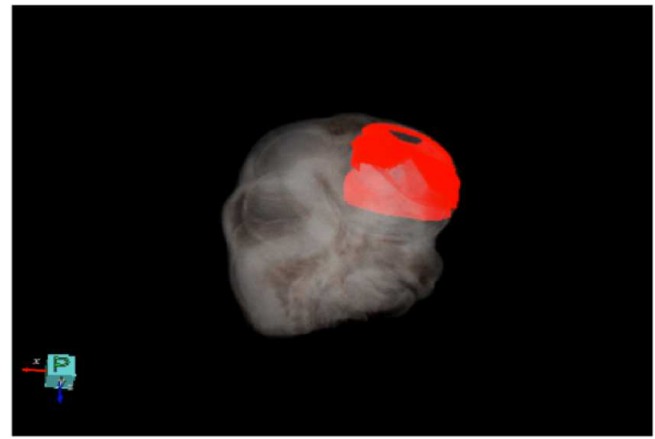}}
\caption{Extrapolating gene expression patterns to form a prediction of the full 3-dimensional expression pattern}
\label{fig:pattern_extrapolation}
\end{center}
\end{figure}

After images resulting from \emph{in-situ} studies are mapped to embryo models, with optionally, missing sections predicted using extrapolation, it is vital good visualisation techniques are available. These techniques should support the general questions researchers ask about gene expression patterns. Such questions include comparisons between the relative spatial positions of patterns of various genes and between the shape of expression patterns in different embryos. We have developed tools that allow researchers to make use of techniques that support answering these type of questions.

\subsection{Human-Mouse Comparison} \label{s:humanmouse}

To understand how genes determine the human developmental process it is necessary to capture data from human embryos. There are many examples, especially in the brain where the pattern of expression in say mouse is significantly different to that in human for the homologous gene. Nevertheless the number of genes that have mouse homologues between say mouse and human  is about 98\% and about 91\% of the human genome lies in conserved syntenic segments (with respect to mouse).
The implication is that the mouse can serve as a good model for understanding mammalian development and direct comparison between mouse and human will be a way to establish probable gene function and gene-network activity in the matching spatio-temporal context. Hypotheses generated in this way of course then need to be checked in the human. It is very important therefore that there is efficient interoperability between the human and mouse data-resources and that data from one can be visualised and analysed in the context of the other.

The key property of the gene-expression data associated with DGEMap is the spatial pattern at a given stage of development. For interoperability we therefore need both spatial and temporal mappings between the human and mouse. This is in addition of course to the ``standard'' bioinformatics mapping of genes and sequences via sequence comparison and homology identification. For this purpose we are therefore developing direct spatial mappings (3D warp transformation) between the standard reference models of the mouse and human databases. These have been tested in prototype and have been integrated in the context of the query and analysis tools available via the DGEMap portal.

\subsection{Data Mining on Gene Expression Patterns} \label{s:datamining}

The increase in efficiency of \emph{in-situ} hybridisation experiments brings with it a significant increase in the amount of data that needs to be analysed. For instance, the Emage Gene Expression Database Repository~\cite{VenkataramanStevenson2008} contains 29560 assays, where each assay can hold up to 24 images. With such a rich resource on data, it is impossible for a researcher to find relationships between data in a structured manner. Instead, we need to support researchers with methods that allow structured searches and which can extract significant relationships from massive amounts of data. Data mining~\cite{TanSteinbach2005} provides several methodologies for extracting knowledge from data by ordering data, building rules to describe data and by ordering these rules according to measured properties such as significance and interestingness. As most of the methodologies are designed for traditional nominal data, these need to be adapted to suit the domain of gene expression patterns. Here we discuss two of those adapted methods.

\begin{figure*}[htp]
\subfigure[Spatial context of a strong association rule \emph{Brap,Zfp354b} $\Rightarrow$ \emph{9830124H08Rik}, which has a support of 6.0\%, a confidence of 97.9\% and a lift of 10.2]{\includegraphics[width=0.33\textwidth]{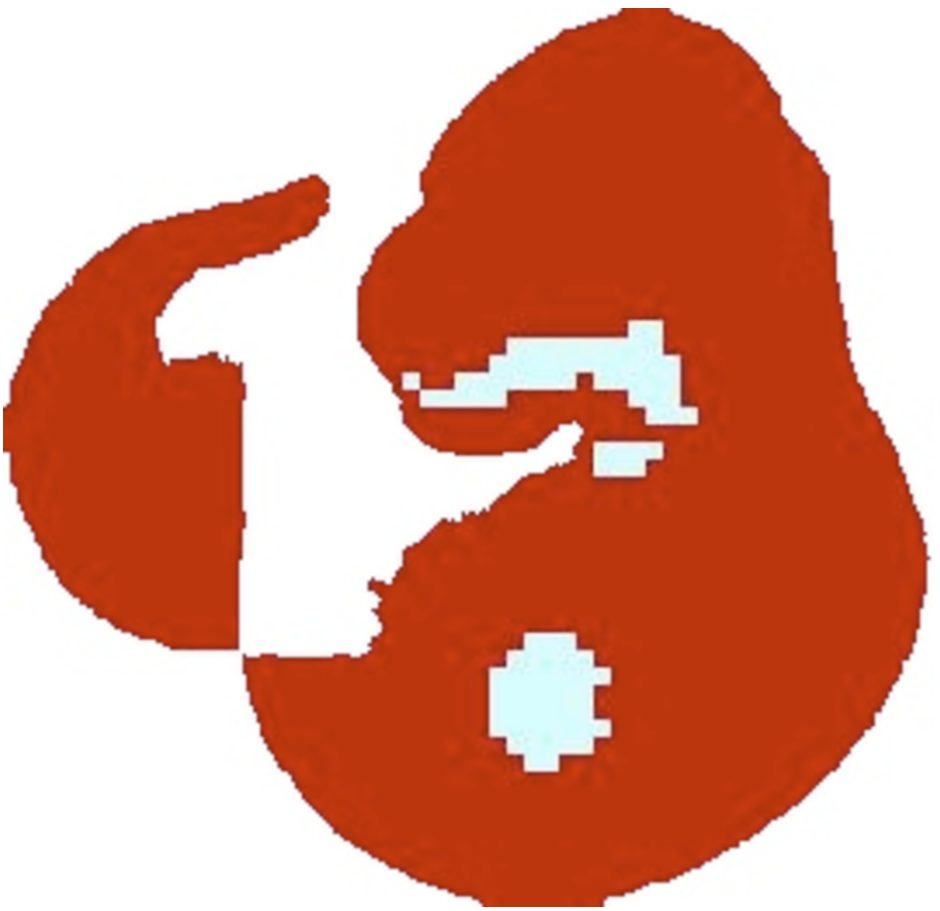}\label{fig:associationrule}}
\hspace{1em}
\subfigure[Left, the target pattern, which consists of strong and moderate expression of the gene Mid1. Right, the best match with a similarity index of 0.753, which uses a specific combination of the spatial gene expression patterns of the genes Rnf34, Pax5 and Anapc11]{\includegraphics[width=0.33\textwidth]{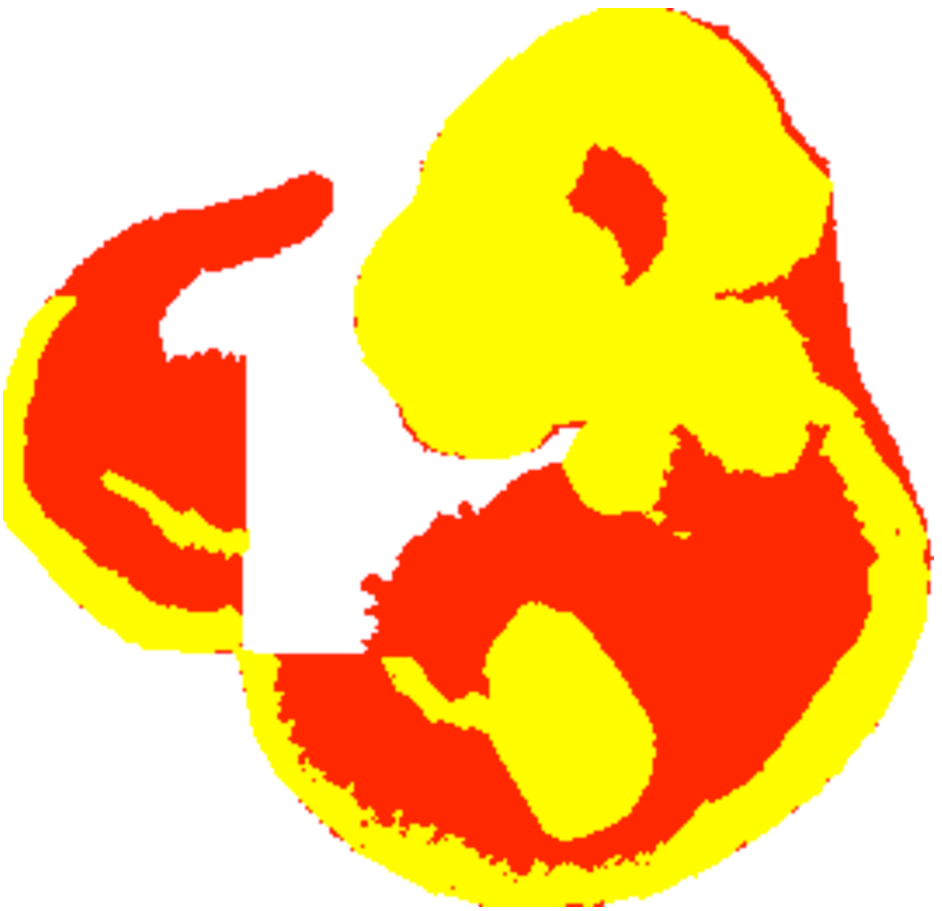}\includegraphics[width=0.33\textwidth]{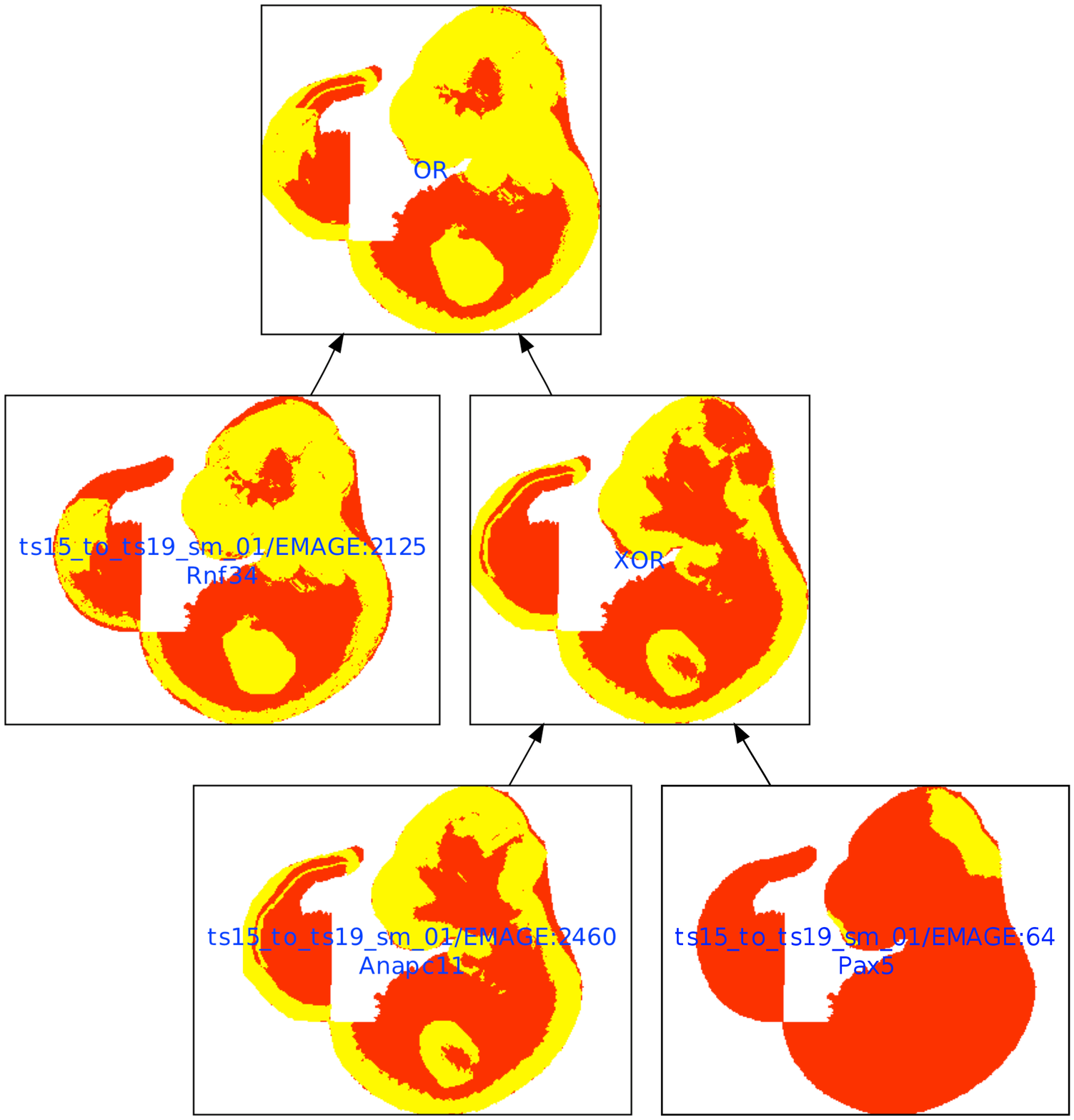}\label{fig:spatialtree}}
\caption{Examples of results from applying novel data mining techniques in the domain of curated spatio-temporal gene expression patterns}
\end{figure*}

We introduce a novel application of mining for association rules in results of \textit{in situ} gene expression studies. In earlier studies in which association rules were applied to gene expression results, these results originated from micro-array experiments, where the aim is to find associations between genes~\cite{BecquetBlachon2002a} in the context of broad tissue types. Here in contrast, we will operate on accurate spatial regions with patterns derived from \textit{in situ} experiments. This type of accurate data enables us to extract two types of interesting association rules, first we can extract the same type of relationships between genes; an example of which is shown in Figure~\ref{fig:associationrule}. However, we can also extract rules expressed in the form of spatial regions, thereby providing knowledge on how areas in an embryo are linked spatially. The only other study in the direction of spatial association rules the authors are aware of, is solely based on synthetic data~\cite{OrdonezOmiecinski1999a}. A more detailed study on our application is available in~\cite{HB2007}.

We have developed a methodology to define a given target pattern by the combination of multiple gene expression patterns via several gene interactions operations. The target pattern can be the expression pattern of a particular gene, a pattern defined by a human, a pattern defined by anatomical components or by any other means of defining spatial area within the context of the model mouse embryo. The method searches for a set of genes and combines their expression patterns using predefined operations to closely match the target pattern, thereby attempting to define this pattern spatially. The objective of this study is to measure the robustness of the methodology and validate the significance of the resulting gene interactions. This is important as much noise exists in the acquisition of the patterns as well as much inaccuracy may exist in the target pattern. An example of a significant result is provided in Figure~\ref{fig:spatialtree} and a more detailed study is available in~\cite{HB2008}.

\section{Related Work}

There are several other projects which propose similar e-Infrastructure to support the work of research scientists in the laboratory, a summary of such systems is presented in~\cite{fgcs}. The most widely known are \emph{myExperiment} (\url{http://www.myexperiment.org/}) a social networking Website for scientists to upload and share scientific workflows and their associated results and \emph{ViroLab}  (\url{http://www.virolab.org/}) a Web-based virtual laboratory framework for infectious diseases. 

Our e-Infrastructure provides a unified solution for human embryo research, accessible through a Web-based portal, supporting the research cycle of a scientist by providing functionality to: share physical lab resources, execute, share and monitor workflows and results, image processing, visualisation and data mining services. In comparison existing solutions focus on a particular aspect of this research cycle, e.g. myExperiment focuses primarily on sharing workflow definitions in order to encourage reuse.  

\section{Discussion}

We have identified some of the key challenges in enabling \emph{in-situ} studies of human embryonic tissue. These challenges include overcoming the problems associated with laboratories that are geographically distributed over many countries; handling of large quantities of experiment and result data from these laboratories; working with human embryonic tissue, which is very a scarce resource; dealing with a relatively small amount of results on human development; and coping with large quantities of results from \emph{in-situ} experiments when comparing with mouse-models. 

The Developmental Gene Expression Map project has performed a three year design study that focused on tackling these challenges in the context of feasibility, ethics, biological aspects and electronic infrastructure. In this paper, we have focussed on the latter, which has resulted in an e-infrastructure consisting of several components. These components, when integrated, address the following important aspects of the challenges. A portal was designed to enable a single point of entry to data and methodologies, which allows researchers from any geographical location to collaborate on human embryonic tissue by facilitating access to the several human embryo resources, to results from experiments, and to advanced methods for processing and visualising these results. New data transportation protocols were investigated to allow processing of large volumes of data over distributed sites using standard workflow tools. To make optimal use of scarce human embryonic tissue, new image and visualisation techniques were developed that allow missing data to be predicted. Spatial and temporal mappings between human and mouse models were established to allow much for a much richer environment for building new hypotheses. Last, to cope with the large volume of data made available through these mappings, novel data mining techniques were applied to synthesise new hypotheses.

Members of the DGEMap project are investigating possible ways of implementing the design on a larger scale over the laboratories in Europe that collect human embryonic tissue as well as those that perform research on human development.

\section{Acknowledgements}The e-infrastructure work package is part of an EU-funded Design Study \emph{Developmental Gene Expression Map} (contract number 011993; \url{http://www.dgemap.org/}), which is led by professor Susan Lindsay at Newcastle University, UK. 

\bibliographystyle{abbrv} \bibliography{dgemap_barker}

\end{document}